## Understanding surface potential dynamics of passivated perovskites via Kelvin Probe Force Microscopy

Rehmat Sood-Goodwin[1,2], Xue-Li Cao[3,4], Benjamin C Kinvig[1], Robert D. J. Oliver[5], Yen-Hung Lin[3,4], Nic Mullin[1] and Alexandra J. Ramadan[1]*

[1]School of Mathematical and Physical Sciences, University of Sheffield, Hicks Building, Hounsfield Road, Sheffield, S3 7RH, UK

[2]Grantham Centre for Sustainable Futures, University of Sheffield, Addison Building, Sheffield, S10 2TN, UK

[3]Department of Electronic and Computer Engineering, The Hong Kong University of Science and Technology, Hong Kong SAR, China

[4]State Key Laboratory of Displays and Opto-Electronics, The Hong Kong University of Science and Technology, Hong Kong SAR, China

[5]School of Chemical, Materials and Biological Engineering, University of Sheffield, Sir Robert Hadfield Building, Mappin Street, Sheffield, S1 3JD, UK

Corresponding author: a.ramadan@sheffield.ac.uk

### Abstract

Molecular passivation has become central to reducing photovoltage losses in metal-halide perovskite solar cells, but its electronic action is still often inferred from device-level metrics rather than directly resolved at the nanoscale. Here, we use amplitude-modulated Kelvin probe force microscopy to examine how [3-(2-aminoethylamino)propyl]trimethoxysilane (AEAPTMS) modifies the surface potential and photovoltage dynamics of mixed-cation, mixed-halide perovskite thin films. AEAPTMS homogenises the dark contact potential difference, narrowing its distribution from ~45.7 to ~14.6 mV without obvious morphological changes. Under illumination, passivated films show a larger steady-state surface photovoltage (SPV) and faster stabilisation, with the SPV increasing from ~345 to ~417 mV and the stabilisation time constant decreasing from ~840 to ~470 s. Wavelength-dependent SPV further indicates reduced sub-bandgap electronic disorder. By separating grain-boundary and grain-interior contributions, we show that AEAPTMS suppresses grain-boundary potential barriers, linking amino-silane passivation to a more homogeneous and stable carrier landscape.

### Introduction

Metal halide perovskites possess excellent optoelectronic properties such as high absorption coefficients[1], long carrier diffusion lengths[2,3], tuneable bandgaps[4,5], and high defect tolerances[1,6]. Whilst this defect tolerance exceeds that of inorganic semiconductors, interfacial defects within optoelectronic devices can still significantly degrade device performance[7–9]. The loss of open-circuit voltage, $V_{OC}$, primarily arises from non-radiative recombination[10–13], which is exacerbated by structural imperfections such as grain boundaries and surface interfaces within polycrystalline perovskite thin films[7,10,11,14,15]. Chemical passivation of defects at the perovskite/charge transport layer interfaces has been widely employed to mitigate their effect on device efficiency[1,13,16–18]. While macroscopic measurements can identify the overall efficiency gains from these passivation molecules, fewer studies have resolved the effects of molecular passivation at the micro- and nanometre scales[8].

Over the last decade, Kelvin Probe Force Microscopy (KPFM) has evolved as a pivotal tool for uncovering these dynamics in perovskites, allowing researchers to track carrier behaviour over temporal[19], spatial[20], and cross-sectional domains[21]. Unlike macroscopic techniques such as photoelectron spectroscopy, KPFM allows the Contact Potential Difference (CPD) to be correlated directly with nanoscale surface topography. Surface photovoltage (SPV) is the change in CPD from dark to under illumination, which can serve as a localised proxy for the Quasi-Fermi Level Splitting (QFLS)[22–24]. This is vital for identifying how local topographical changes or shifts in surface potential cause bottlenecks in carrier extraction[8,11]. Surface and interfacial defects act as electronic shunts that narrow the Quasi-Fermi Level Splitting (QFLS), which dictates the upper limit of the $V_{OC}$[8,25].

In this work we use Amplitude Modulation Kelvin Probe Force Microscopy (AM-KPFM) to map topography and CPD of 1.60 eV perovskite thin films with one promising amino-silane passivation molecule[1,13,16]. We find that the passivation reduces the heterogeneity in the CPD by a factor of 3. We measure the evolution of the SPV over several hours under controlled illumination and find that passivated films give larger steady-state SPV values (21% increase) and observe a 44% decrease in the time constant, indicating faster responses to illumination. We also utilise wavelength-dependent KPFM measurements to decouple surface and bulk dynamics within the perovskite films. Finally, we analyse the CPD as a function of distance from grain-grain interfaces and show the significant impact of grain boundaries on surface carrier distributions. The analyses developed here are broadly applicable to other perovskite systems and passivation chemistries, and we provide open-source analysis code for both temporal and spatial KPFM measurements.

**Results and discussion**

Kelvin Probe Force Microscopy (KPFM) measures the CPD between the probe and the sample surface[2–4]:

$$CPD = \frac{(\Phi_{probe}\text{-}\Phi_{sample})}{e} \quad (1)$$

Where Φ is the work function of the probe or sample, and e is the elementary charge. The CPD is suggestive of a representative distribution of features such as defect densities across the film surface at the nanoscale[26–28]. Furthermore, we can measure the change in CPD upon illumination relative to the CPD measured in the dark (SPV)[23,25,27,29]:

$$SPV = CPD_{light} - CPD_{dark} \quad (2)$$

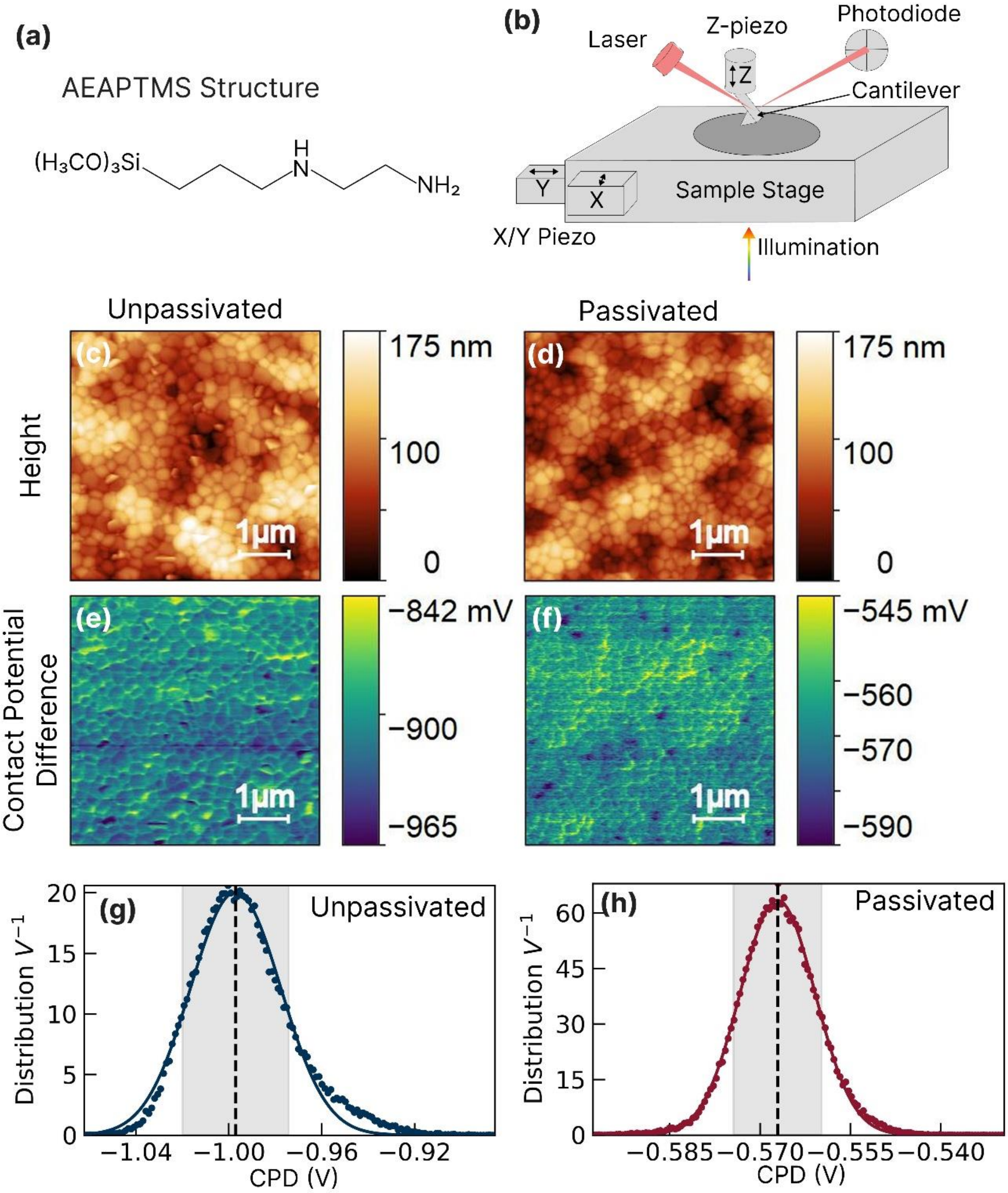

*Figure 1: (a) Chemical structure of [3-(2-aminoethylamino)propyl]trimethoxysilane (AEAPTMS), (b) Schematic of Kelvin probe force microscopy (KPFM) set-up with arrow indicating illumination direction. KPFM images of $FA_{0.83}Cs_{0.17}Pb(I_{0.9}Br_{0.1})_3$ on ITO glass showing the height (c, d) and contact potential difference (CPD) (e, f). (g, h) Collated pixel distributions of the CPD images in (e) and (f) (respectively), where the solid line is the fitted distribution, shaded region is the full-width half-maximum of the distribution, and the dashed line indicates the center of the fitted curve for representative unpassivated and passivated films respectively.*

SPV is a localised measurement of a shift in band position under illumination so is indicative of the $V_{OC}$[15]. In the context of KPFM, this allows us to map the relative changes in the surface potential across the film surface, therefore distinguishing specific morphological features and their impact on carrier dynamics[5–7].

We examined the impact of amino-silane molecules (i.e. [3-(2-aminoethylamino)propyl]trimethoxysilane (AEAPTMS), Fig 1a) on the surface potential and morphological properties of $FA_{0.83}Cs_{0.17}Pb(I_{0.9}Br_{0.1})_3$ perovskite thin films (~1.60 eV), using dual-pass amplitude-modulated Kelvin probe force microscopy (AM-KPFM). Our KPFM set up is illustrated in Fig 1b, notably illumination is applied to the samples through the glass side, allowing for simultaneous mapping of the top perovskite surface, to mimic working conditions. Full experimental protocols for these images can be seen in Notes S1 and S2, and analysis protocols Note S3.

We show representative height (Fig 1c, d) and CPD (Fig 1e, f) images and corresponding CPD distributions (Fig 1g, h) under dark conditions for the unpassivated and passivated perovskite films respectively. Both perovskite systems are composed of ~50-500 nm grains with no apparent changes to the morphology of the perovskite observed due to the passivation molecules (Fig 1c-d).

The corresponding CPD images are shown in Fig 1e-f, where the electronic impact of this passivation on electronic heterogeneity is visible. We quantify this impact in Fig 1g-h, which shows the distribution in CPD values integrated across the whole images, and a FWHM of the collated pixel distributions is taken as a quantitative indication of spatial heterogeneity. The unpassivated perovskite (Fig 1g) shows considerable spatial heterogeneity of CPD values (~45.7 +/- 0.2 mV) across the image with a visible surface potential contrast between the grain boundaries and grain interiors. Conversely, the passivated perovskite (Fig 1h) shows a smaller distribution of CPD values (~14.64 +/- 0.02 mV). The increased homogeneity of the CPD maps with passivation suggests that the AEAPTMS molecules effectively passivate surface defect states, leading to a more uniform potential landscape.

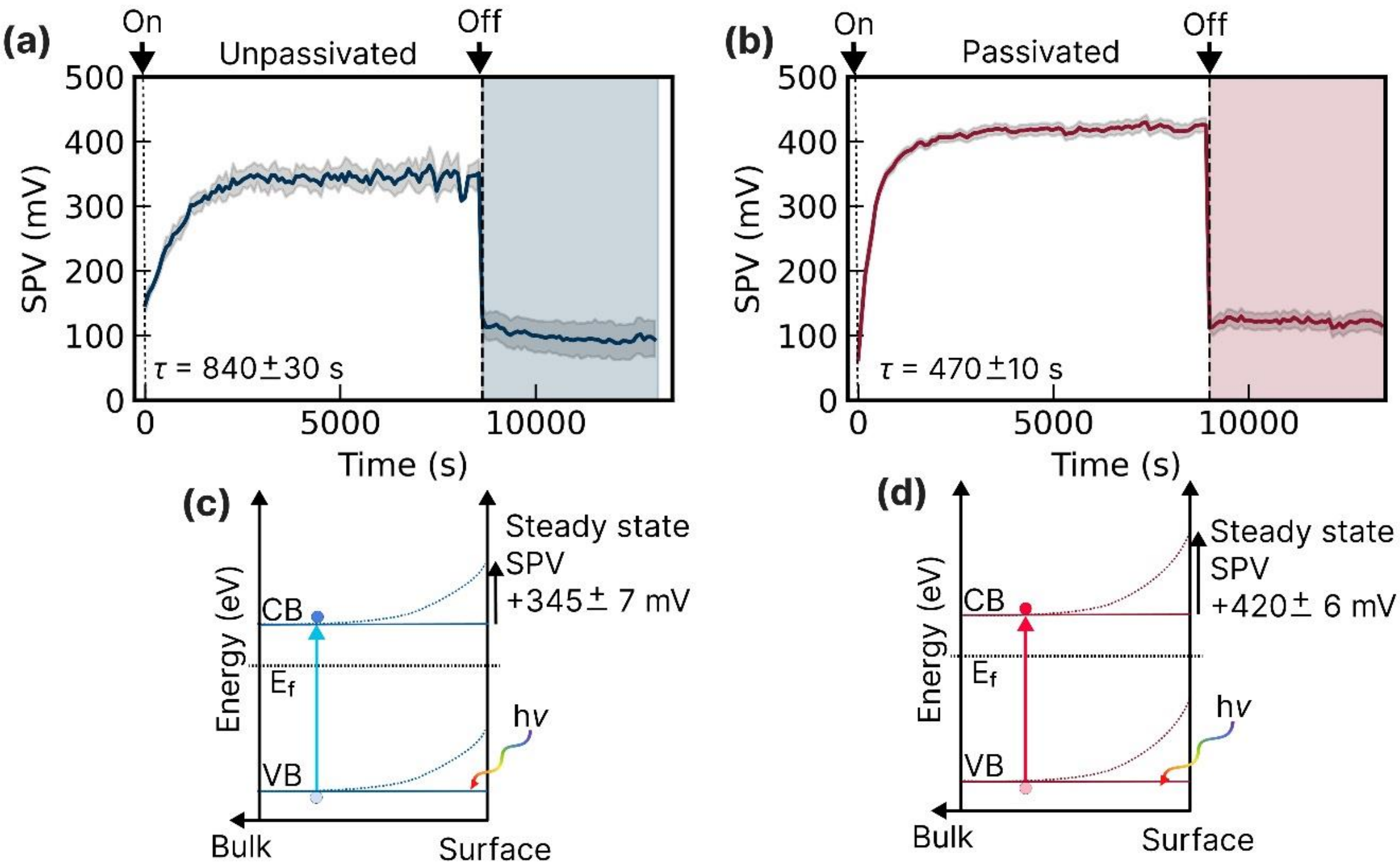

*Figure 2: Temporal surface photovoltage (SPV) characteristics for (a) unpassivated and (b) passivated perovskite thin films. Dashed lines indicate points at which illumination is turned on or off. Time constant, $\tau$, is related to the time taken to reach a steady state SPV. Grey shaded areas correspond to spread (FWHM) of each image, and solid lines indicate the peak SPV from fitting (See Fig S1a-d). Blue and pink shaded regions indicate post-illumination region for unpassivated and passivated films respectively. (c, d) Schematic of band diagram for unpassivated and passivated samples respectively. Dashed arrow indicates carrier excitation in bulk under illumination. Solid arrow denotes a positive steady state SPV indicating hole accumulation at the probe-sample surface, and is calculated using exponential fitting shown in Fig S1e-f.*

The shift in the peak of the CPD is not necessarily indicative of physical processes associated with the perovskite itself, but likely due to factors such as uncalibrated.

The influence of passivation on SPV was investigated by measuring continuously both under illumination with broad spectrum white light of ~1 sun intensity and post-illumination in dark conditions. These measurements were performed by undertaking continuous KPFM scans every 90s (2.5x5 $\mu$m/32x64 pixels), where the illumination is turned on at the start of the first ‘light on’ scan, and off at the start of the first “light off” scan (See Note S1 and S4 for experimental details, and Fig S1 for analysis processes). The resulting SPV response during and post-illumination, taken from distributions of the full images (See Fig S1a-d), exhibit three characteristic regions of interest: (i) an initial rapid rise in SPV upon illumination, slowing to reach, (ii) ”steady-state” SPV under illumination, followed by (iii) a rapid reduction in SPV once illumination is switched off and (iv) smaller changes to the SPV over minutes to hours after the illumination is switched off.

To compare the two illumination curves, we fitted the illumination regions of Fig 2a and b, shown in Fig S1e-f, using an exponential curve (Note S5, Fig S1). The time taken for the SPV to rise within a factor of 1/e of its steady state is the time constant, τ. We use τ as an indication of the time taken for the film to begin stabilising, where a smaller τ indicates a faster time to reach stabilisation. Notably, τ for the unpassivated system was a factor of 1.8x larger than that of the passivated film (840 ± 30 s unpassivated vs. 470 ± 10 s passivated). This accelerated stabilisation is suggestive of the AEAPTMS successfully passivating slow-acting defect states, which commonly govern long-lived transients observed in perovskites[8].

We determine a steady state SPV value as A (via the same fitting used to calculate τ), plus C which is taken from the SPV of the first illuminated image. The steady state SPV magnitude has increased under illumination by a factor of 1.23 (fitting shown in Fig S1 (~345 ± 7 mV unpassivated (Fig S1e) vs ~417 ± 6 mV passivated (Fig S1f)). The enhancement indicates an increase in carrier densities at the surface, which is consistent with a higher $V_{OC}$ via increased carrier separation[22–24]. Furthermore, the sign of the SPV being positive indicates upward band bending at the perovskite-probe interface[10], as illustrated in Fig 2c and 2d. The schematics shown assume that we can only determine the direction of band bending relative to the original band position[30]. Thus, the increased magnitude of the steady state SPV in passivated perovskite films reflects an enhancement of hole accumulation at the surface[8,9], due to more efficient charge separation. The increase in the SPV can be correlated to $V_{OC}$[20,22–24,34], which can be correlated to the increased device performance seen in work from previous groups[1,13,16]. Post-illumination, initial recovery occurs faster than can be resolved in the experiments. The behaviour over minutes to hours post-illumination is indicative of slow-moving carriers diffusing away from the surface and into the bulk. For unpassivated samples there is a slower recovery period in comparison to the passivated samples, which further underscores the electronic stabilisation afforded by AEAPTMS passivation. The persistent SPV post-illumination is also an indication that defect states have been filled and remain so once incident illumination is removed. The increase in SPV (or related metrics) is commonly observed through other techniques[32–34]. This is an indication that there would be improved performance when cycling within illumination, which has been shown in a variety of different works[35–37].

While white-light illumination provides a holistic view of device potential, wavelength-dependent temporal SPV is employed to further investigate the electronic contributions from surface and bulk regions. By selecting wavelengths both above and below bandgap

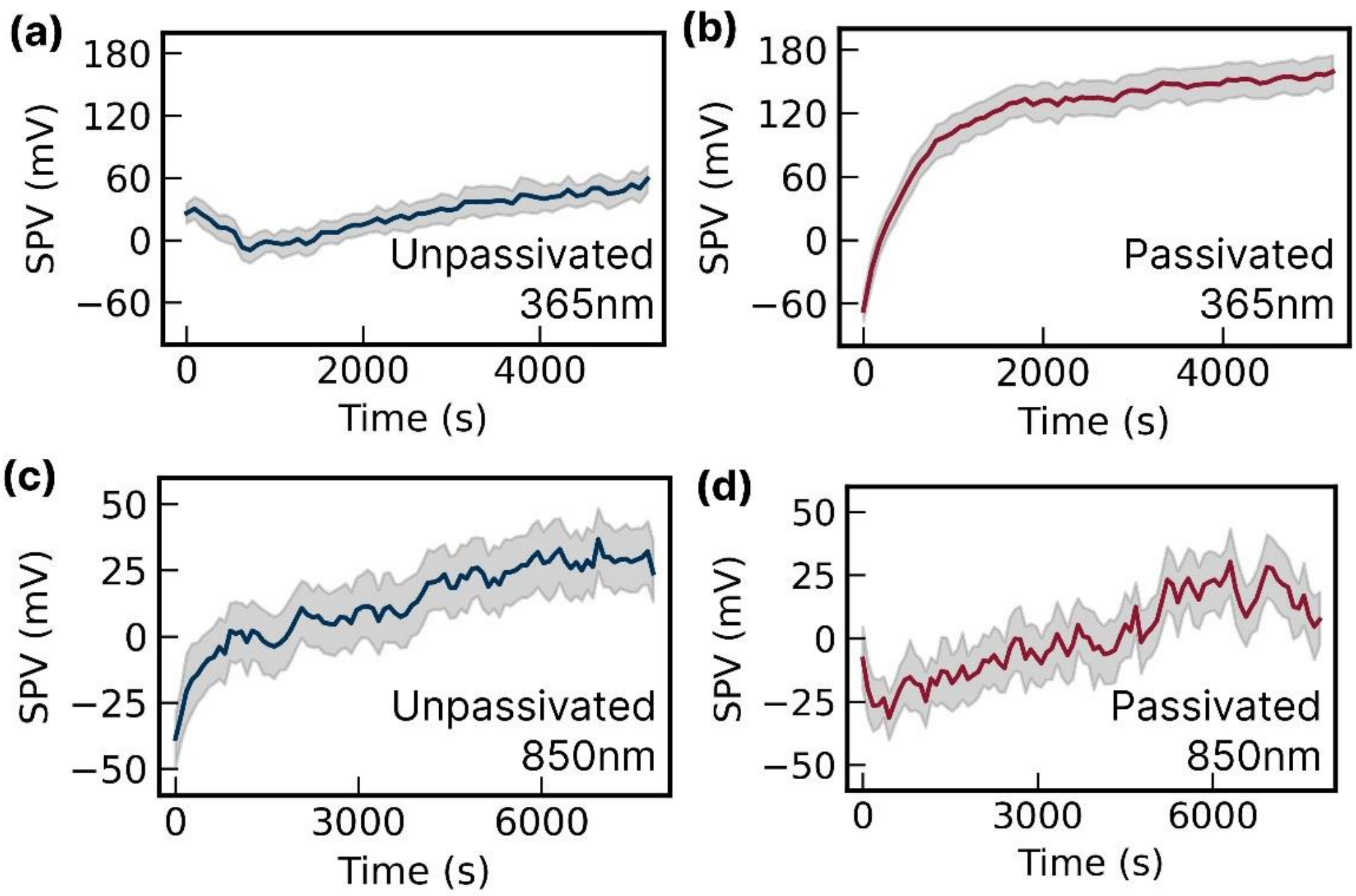

*Figure 3: Temporal surface photovoltage (SPV) measurements under continuous illumination at (a, b) 365 nm and (c, d) 850 nm for unpassivated and passivated perovskite films respectively. Distributions of SPV from each image are fitted and the peak centre taken from multi-gaussian fitting is plotted as the solid line, and grey shaded region indicates spread (FWHM) of images (See Fig S1a-d)*

($E_g$=1.60eV), we can distinguish between surface and bulk electrical contributions and the impact of the AEAPTMS passivation on these dynamics[38–40].

The temporal SPV response is shown in Fig 3a (unpassivated) and Fig 3b (passivated) under 365 nm illumination, allowing us to probe the immediate surface region of the perovskite film[38,41]. Due to illumination being applied to the glass side of our thin films, charge generation at the lower surface is substantial in both unpassivated and passivated films, due to the high absorption coefficients at this wavelength[1,3,4,30]. However, as the SPV is being measured at the top surface of the films, the top interface will play a large role in the measured SPV[30]. The unpassivated (Fig 3a) film demonstrates a lower SPV response, with an initial drop in SPV, followed by a slow, constrained increase in SPV over time. Conversely, the passivated film (Fig 3b), exhibits a steady rise in SPV over time, like the behaviour under white light, towards a much higher steady state SPV. This behaviour suggests that the AEAPTMS is passivating recombination centres at the perovskite surface, leading to more efficient charge separation[41] at the perovskite-probe interface, which is consistent with the increased SPV and reduction of the time constant observed under white light illumination.

Further investigation using 850 nm sub-bandgap illumination is presented in Fig 3c (unpassivated) and Fig 3d (passivated). In an ideal, defect-free 1.60 eV perovskite, there would be little to no change to the temporal SPV, with changes being on a scale that is not discernible from the inherent changes that occur due to KPFM (see SI Fig 3 for more detail). The unpassivated perovskite film response (Fig 3c) exhibits a measurable increase in SPV, but it is not at the same scale as under 365 nm illumination (Fig 3a). In contrast, the passivated sample (Fig 3d) exhibits a lower response, on an order of magnitude that can be compared to dark temporal KPFM (Fig S2). The presence of an SPV response for unpassivated films indicates sub-bandgap absorption, therefore higher SPV at 850 nm is an indication of increased electronic disorder, likely due to the presence of increased defect states compared to passivated samples[42,43]. This is an indication that although the AEAPTMS does not completely nullify the SPV response at 850 nm, it still has a significant impact on the electronic disorder of the perovskite.

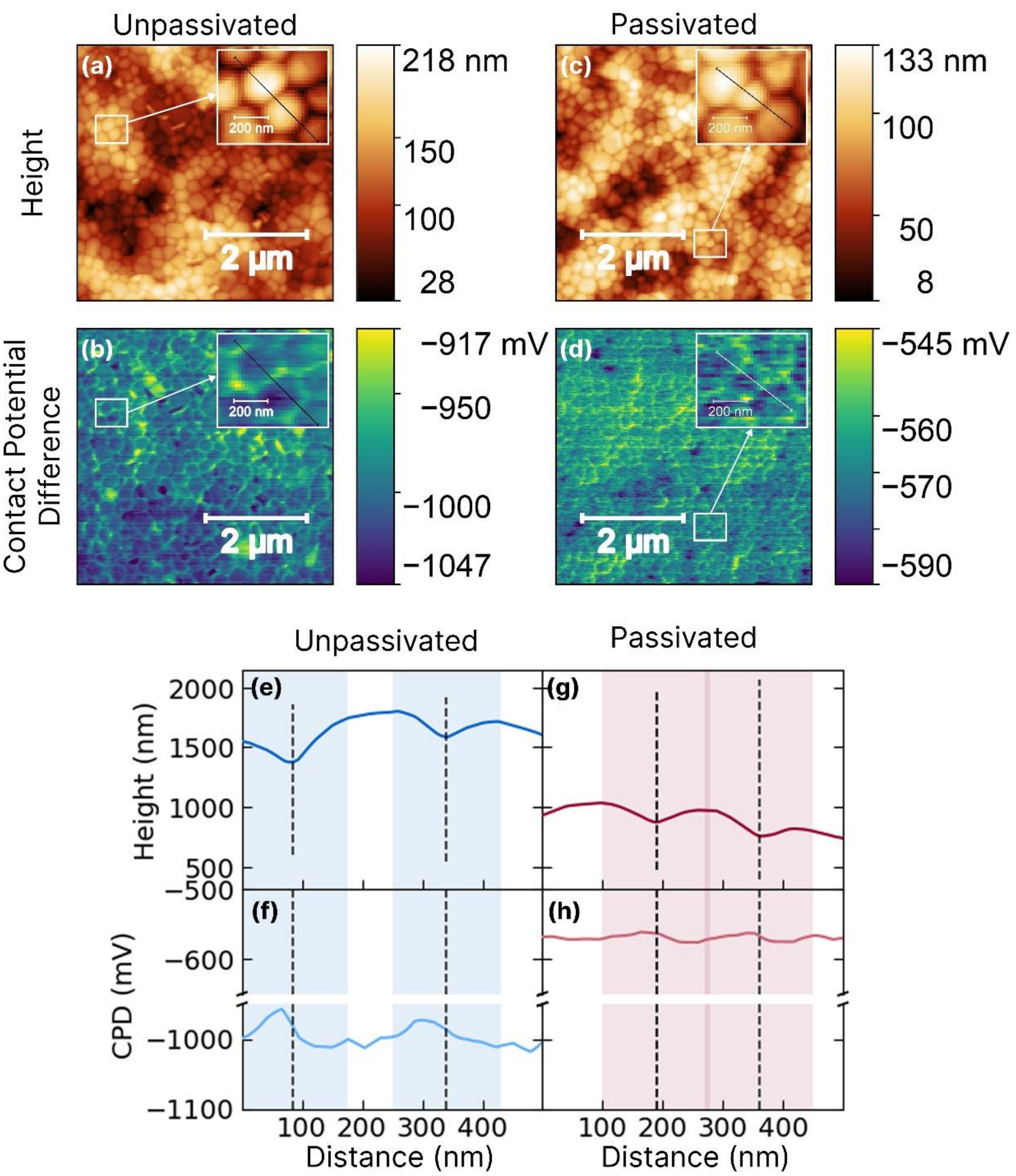

*Figure 4: Height and contact potential difference (CPD) with insets indicating line profile regions for unpassivated (a. b) and passivated (c, d). Line profiles of height and CPD respectively (e, f) unpassivated and (g, h) passivated samples respectively. Black dashed line represents grain boundary (based on minima of height line profile) and shaded regions indicate width of maximum mask region for Figure 5.*

The SPV measurements from Fig 2 and Fig 3 represent how the surface of the perovskite responds under and after illumination, however, this data is from whole images, and the images are clearly heterogeneous, as can be seen in the grey regions of Fig 2a, b and Fig 3, as well as in the images shown in Fig 1. To decouple the influence of morphological heterogeneities from the averaged CPD/SPV, specifically at grain boundaries (GB), from grain interiors (GI), we look at a line profile of height and the respective CPD.

The line profiles are shown in the inset of each image, for height and corresponding CPD images of the unpassivated (Fig 4a-b) and passivated (Fig 4c-d), and plotted in Fig 4e-h, with unpassivated (4e, f), and passivated (4g, h), height profile (4e, g) and corresponding CPD profile (4f, h) shown. The grain boundaries have a clear impact on the unpassivated CPD seen in Fig 4f. The increase in magnitude of CPD at grain boundaries is an indication that there is an accumulation of carriers at the grain boundary, which has been seen in many instances of perovskite KPFM[44,45]. For the passivated film (Fig 4g, h), an increase in CPD is still observed at GBs, but the increase is significantly smaller than that of the unpassivated films. While the cross sections in Fig 4 provide representative examples that provide clear indication of the impact of passivation on GBs, they only utilise a small fraction of the information within the images. Since is it not a straightforward process to collate multiple cross sections, and obtain statistically meaningful data, we have developed an alternative approach, shown in Fig 5a. Briefly, we create a mask from the topographic image channels by tracing the GBs. We then use these masks to segregate the CPD data into ‘grain boundaries’ (masked region) and ‘grain interiors’ (unmasked region), allowing the distributions of CPD values to be compared. (See Note S6 for further experimental and

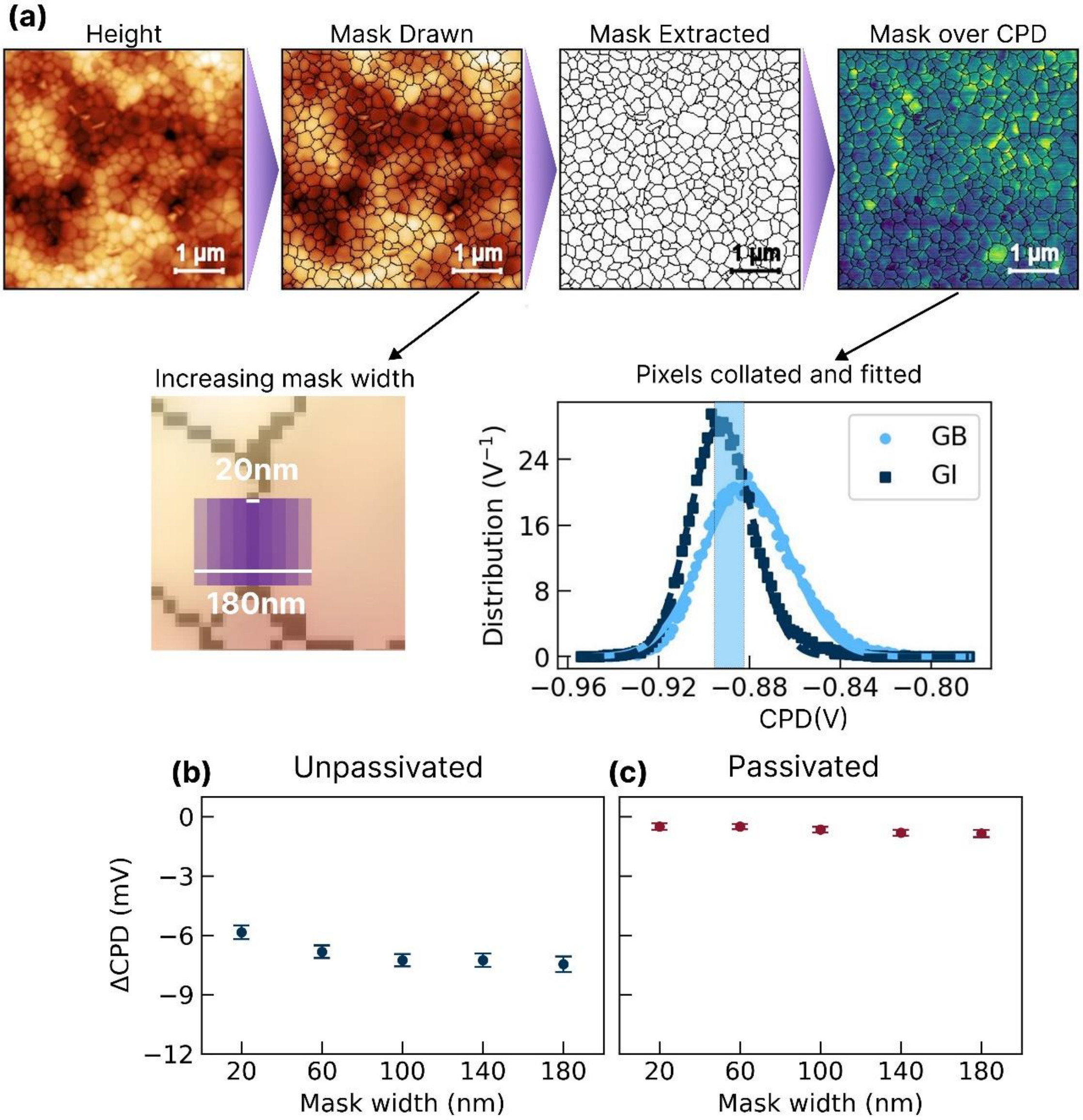

*Figure 5: (a) Demonstration of grain analysis pipeline, further information can be obtained in Supplementary Information. Masks are drawn using height channels and extracted before being distributed onto contact potential difference (CPD) images. Pixel distributions are collated for grain interior (GI) and boundary (GB). Masked region (black) is nominally the GB distribution, and unmasked regions (white) are nominally GI. By increasing mask width GB distribution is made larger, GI distribution is made smaller, see SI fig 4. ΔCPD is calculated using $CPD_{GI} - CPD_{GB}$, indicated with shaded region using the peak of each peak from the respective gaussian fits. (b, c) Scatter plots for unpassivated and passivated films respectively showing ΔCPD ($CPD_{GI} - CPD_{GI}$) dependence with increasing mask widths. Error bars have been calculated from fitting. See Fig S4a, b for scaled plots of unpassivated and passivated samples.*

analysis detail). This method also partially removes the impact of factors such as GB depth/shape by averaging across all grain boundaries within an image. By incrementally increasing mask width we are also able to incorporate most <180 nm grains within the masked region, as smaller grains are commonly associated with lower device performance[46]. Fig 5b and Fig 5c present the resulting ΔCPD values between the grain interiors and boundaries ($\Delta CPD=CPD_{GI}-CPD_{GB}$), for an unpassivated and passivated film respectively. We can use ΔCPD as a measure of the magnitude and slope of the potential barrier at grain boundaries

Firstly, the polarity of ΔCPD is negative for both the unpassivated and passivated films. The polarity of ΔCPD is an indication of which region of the film is dominating the SPV. If it is positive, this is a key indication of which region of the film is dominating the SPV. In this case, for the unpassivated samples (Fig 5b), ΔCPD is significantly more negative across all mask widths compared to the passivated (Fig 5c), therefore the passivation is decreasing the magnitude of carrier accumulation at the grain boundary. This also provides some reasoning behind the reduction in spread of CPD seen in the passivated distribution (Fig 1h) vs. the spread of the unpassivated film (Fig 1g). Thus we are able to conclude from the polarity and magnitude of ΔCPD that the passivation is actively mitigating the effects of carrier accumulation at grain boundaries and improving carrier transport across the film[47–49].

With respect to the dependence of ΔCPD on mask width, we observe a higher dependence for unpassivated films (Fig 5b), compared to the passivated (Fig 5c). However, dependence on mask width can vary depending on image quality, due to factors such as noise, drift, etc. This is more easily visualised in Fig 4e-h, where the grain boundary from the height is offset from the peak CPD. We demonstrate the significant variation in mask width dependence using four images for unpassivated (Fig S4a-d) and passivated (Fig S4e-h) films.

Despite this, the variation of ΔCPD across all images analysed is significantly smaller for passivated than for unpassivated films, suggestive of a decrease in the potential barrier. Thus the minimal dependence of ΔCPD on mask width with passivation indicates the reduction of the potential barrier as $\Delta CPD \rightarrow 0$, further demonstrating the AEAPTMS is passivating GB-related defects[47,50]. By neutralising the lateral heterogeneity associated with grain-to-grain interfaces, the AEAPTMS is decreasing band-bending at the grain boundaries, thereby increasing the film's capability for carrier extraction, and reducing the impact of localised non-radiative recombination[48]. We further affirm that passivation is decreasing grain boundary related defects[49,51,52] using illumination measurements (See Note S7 for experimental details), where we see a larger change in ΔCPD for unpassivated films when the sample is illuminated, compared to the passivated films (Fig S5). This confirms that AEAPTMS is homogenising the film surface, therefore allowing for increased carrier extraction and a reduction in carrier accumulation[53–55].

Collectively, the various KPFM observations build a comprehensive overview of the impact of AEAPTMS on the surface photovoltage stabilisation. Through the interleaving of both wavelength and morphology dependent KPFM measurements, we have been able to decouple surface and bulk contributions to the surface photovoltage and isolate the specific impact of AEAPTMS passivation with respect to morphological features. Suppression of sub-bandgap response under 850nm illumination indicates a reduction in the electronic disorder within the thin films[43]. Furthermore, illumination above bandgap at 365nm and with white light illumination indicate a significant reduction in surface defects, as well as demonstrating the mitigation of light-induced long-term transients with passivation[42,43].

Crucially, the quantitative analysis of morphological features and their relation to CPD reveals the spatial influence of AEAPTMS, demonstrating that it fundamentally homogenises the electronic landscape[1]. The ΔCPD measurements serve as indicators of an increase in homogeneity of, and reduction of band bending at grain boundaries that usually inhibit lateral charge transport, and the changes in ΔCPD under illumination support the conclusion that AEAPTMS passivation has a measurable impact on defects at grain boundaries, by reducing charge accumulation[47,55]. Furthermore, residual, stable SPV post-illumination in Fig 2a-b provides robust evidence that the passivation effectively ‘plugs’ carrier leakage pathways and slow transients; and allows the films to preserve a wider baseline quasi-Fermi level splitting[8].

**Conclusion**

In this work we used AM-KPFM to resolve the nanoscale electronic landscape and carrier dynamics of $FA_{0.83}Cs_{0.17}Pb(I_{0.9}Br_{0.1})_3$ films passivated with AEAPTMS. We show that this interfacial amino-silane passivation homogenises the surface potential, narrowing the measured spatial CPD distribution from ~45.7 mV to ~14.64 mV. Temporal and wavelength-dependent SPV measurements reveal that passivated films exhibit a 1.8-fold acceleration in SPV stabilisation under illumination in addition to an enhanced steady-state SPV (~417 mV vs. ~345 mV), suggesting suppressed non-radiative recombination and increased quasi-Fermi level splitting. We developed an image analysis framework to enable quantification of CPD changes at grain boundaries. Using our approach, we show the AEAPTMS results in a significant mitigation of ΔCPD at grain boundaries, supporting the conclusion that GB-specific defect states have been neutralised and the reduction of local potential barriers to lateral charge transport. Collectively, these nanoscale insights directly correlate targeted interfacial passivation with minimized carrier leakage and reduced electronic disorder.

**Methods**

**Perovskite thin film fabrication**

$PbI_2$ (99.99%, TCI), $PbBr_2$ (>98.0%, TCI), FAI (>99.99%, Dyenamo) and CsI (99.9%, Alfa Aesar) were used as received; all remaining reagents and solvents were sourced from Sigma-Aldrich. Precursor solutions targeting the $FA_{0.83}Cs_{0.17}Pb(I_{0.9}Br_{0.1})_3$ composition were prepared by dissolving the four salts in stoichiometric proportions in a 4:1 (v/v) mixture of N,N-dimethylformamide (DMF) and dimethyl sulfoxide (DMSO) to give a 1.45 M solution, which was stirred overnight under nitrogen before use.

ITO-coated glass substrates underwent ultrasonic cleaning in successive baths of Decon 90 solution (2% v/v in deionised water), deionised water, acetone and isopropanol for approximately 10 min each, followed by nitrogen drying and a 20 min UV-ozone treatment. Perovskite layers were spin-coated under nitrogen at 1,000 rpm (5 s, 5 s ramp) then 5,000 rpm (30 s, 5 s ramp); at 35 s, 200 µL of anisole was dispensed as an antisolvent quench, after which the films were annealed at 100 °C for 50 min. Surface passivation was carried out by vapour-phase exposure to AEAPTMS: the molecule was placed in a glass petri dish (Ø 100 × 20 mm), heated to 100 °C until fully vaporised, and the perovskite film was sealed inside the dish for 100 ± 20 s. Further experimental detail for film fabrication can be found in Lin et al (2024)[6].

**Kelvin Probe Force Microscopy**

All Kelvin Probe Force Microscopy (KPFM) is conducted on an MFP-3D microscope (Asylum Research, Oxford Instruments), using Multi75-G probes (nominal spring constant 3 N/m, nominal resonant frequency 75 kHz) (BudgetSensors). The AFM is mounted on an inverted optical microscope (IX-71, Olympus), which is used for controlled illumination of the sample from underneath. Further details on illumination sources can be seen in SI note 8, and the spectra of the broad-spectrum white lamp can be seen in Fig S6.

Reference images and grain analysis, images taken are either square (5x5$\mu$m; 256x256 pixels) or rectangular (5x2.5$\mu$m; 256x128 pixels). Rectangular scans are specific to SPV measurements to decrease gradual changes along the slow scan axis during illumination and recovery. Full details of all experimental protocols, illumination characterisation and data processing are given in Supporting Information. General imaging parameters are given in SI table S1. The code corresponding to the analysis pipeline discussed in Figure 5 and used in analysis throughout is available at https://github.com/RamadanLab/KPFM_processing.

**Photoluminescence spectroscopy**

Photoluminescence spectroscopy was used to obtain spectra of both unpassivated and passivated films, and measurements were taken for the perovskite and glass side of each sample (See Fig S6a,b). Samples were placed in an integrating sphere (819C-IS-5.3, Newport) and illuminated with a 405 nm, 50mW Low-Cost Turnkey Laser (Edmund Optics), with a spot diameter of ~0.55mm$^2$, with a power of ~46.11mW/cm$^2$. Powers were calibrated using a HL-3P-INT-CAL lamp (Ocean Sight). Spectra were collected and plotting was done in Origin 2024b.

**UV-Visible absorbance spectroscopy**

UV-Visible absorbance was taken using a Fluromax-4 fluorometer (Horiba) over a spectral range of 600 to 900 nm, and absorbance was calculated using the equation:

$$A = -\log_{10}(I/I_0)$$

Spectra shown in Fig S6f,d. Data was processed and plotted in Origin 2024b.

**Acknowledgements**

The authors thank Shing Wan for his support with streamlining of analysis code for open-source use, and Simon Dixon from the Physics Workshop in Sheffield for aluminium sample holders. The authors acknowledge funding from the Engineering and Physical Sciences Research Council (EPSRC) (EP/X039285/1), X. R.S-G. acknowledges the Grantham Centre for Sustainable Futures for funding. Y.-H.L. and X.-L.C. acknowledge support from the Hong Kong Research Grant Council Early Career Scheme (no. 26210623) and the State Key Laboratory of Displays and Opto-Electronics.

**Author Contributions**

Rehmat Sood-Goodwin: conceptualization, methodology, investigation, formal analysis, writing – original draft, review and editing.
Xueli Cao: investigation, writing- review and editing.
Benjamin C. Kinvig: software, formal analysis, writing – review and editing
Robert D. J. Oliver: conceptualization, supervision, writing – review and editing
Yen-Hung Lin: conceptualization, funding acquisition, supervision, writing – review and editing
Nic Mullin: conceptualization, methodology, supervision, writing – original draft, review and editing.
Alexandra J. Ramadan: conceptualization, funding acquisition, methodology, supervision, writing – original draft, review and editing.

**Conflict of interest**

The authors declare no competing financial interest.